\documentclass[twocolumn,showpacs,aps,
preprintnumbers,amsmath,amssymb,superscriptaddress]{revtex4}
\usepackage{comment}
\usepackage{epsfig}
\usepackage{amsmath}
\usepackage{ulem}
\usepackage{graphicx}
\usepackage{dcolumn}
\usepackage{bm}

\newcommand{\al}{$\alpha$}

\newcommand{\snt}{$^{120}$Sn}
\newcommand{\hes}{$^6$He}
\newcommand{\raa}{($\alpha$,$\alpha$)}
\newcommand{\rhes}{($^6$He,$^6$He)}

\newcommand{\rpg}{(p,$\gamma$)}

\newcommand{\str}{$\sigma_{\rm{reac}}$}

\begin{document}

\title{Comparison of $^{120}$Sn($^6$He,$^6$He)$^{120}$Sn and
  $^{120}$Sn($\alpha$,$\alpha$)$^{120}$Sn elastic scattering and signatures of
  the $^6$He neutron halo in the optical potential
}

\author{P.\,Mohr}
\affiliation{
Strahlentherapie, Diakonie-Klinikum, D-74523 Schw\"abisch Hall, Germany}
\affiliation{
Institute of Nuclear Research (ATOMKI), H-4001 Debrecen, Hungary}
\date{\today}
\author{P. N. de Faria}
\affiliation{{\it Instituto de Fisica-Universidade de S\~ao Paulo, C.P.66318,05389-970 S\~ao Paulo, Brazil}}
\author{ R. Lichtenth\"aler}
\affiliation{{\it Instituto de Fisica-Universidade de S\~ao Paulo, C.P.66318,05389-970 S\~ao Paulo, Brazil}}
\author{ K. C. C. Pires }
\affiliation{{\it Instituto de Fisica-Universidade de S\~ao Paulo, C.P.66318,05389-970 S\~ao Paulo, Brazil}}
\author{  V. Guimar\~aes }
\affiliation{{\it Instituto de Fisica-Universidade de S\~ao Paulo, C.P.66318,05389-970 S\~ao Paulo, Brazil}}
\author{ A. L\'epine-Szily }
\affiliation{{\it Instituto de Fisica-Universidade de S\~ao Paulo, C.P.66318,05389-970 S\~ao Paulo, Brazil}}
\author{ D. R. Mendes Junior }
\affiliation{{\it Instituto de Fisica-Universidade de S\~ao Paulo, C.P.66318,05389-970 S\~ao Paulo, Brazil}}
\author{ A. Arazi }
\affiliation{ {\it Laboratorio Tandar, Departamento de Fisica, Comisi\'on Nacional de Energ\'ia}}
\author{  A. Barioni}
\affiliation{{\it Instituto de Fisica-Universidade de S\~ao Paulo, C.P.66318,05389-970 S\~ao Paulo, Brazil}}
\author{V. Morcelle}
\affiliation{{\it Instituto de Fisica-Universidade de S\~ao Paulo, C.P.66318,05389-970 S\~ao Paulo, Brazil}}
\author{ M.C. Morais}
\affiliation{{\it Instituto de Fisica-Universidade de S\~ao Paulo, C.P.66318,05389-970 S\~ao Paulo, Brazil}}

\begin{abstract}
Cross sections of $^{120}$Sn($\alpha$,$\alpha$)$^{120}$Sn elastic scattering
have been extracted from the $\alpha$ particle beam contamination of a recent
$^{120}$Sn($^6$He,$^6$He)$^{120}$Sn experiment. Both reactions are analyzed
using systematic double folding potentials in the real part and smoothly
varying Woods-Saxon potentials in the imaginary part. The potential extracted
from the $^{120}$Sn($^6$He,$^6$He)$^{120}$Sn data may be
used as the basis for the construction of a simple global $^6$He optical
potential. The comparison of the $^6$He and $\alpha$ data shows that the halo
nature of the $^6$He nucleus leads to a clear signature in the reflexion
coefficients $\eta_L$: the relevant angular momenta $L$ with $\eta_L \gg 0$
and $\eta_L \ll 1$ are shifted to larger
$L$ with a broader distribution. This signature is not present in the
$\alpha$ scattering data and can thus be used as a new criterion for the
definition of a halo nucleus.
\end{abstract}


\maketitle

\section{Introduction}
\label{sec:intro}
In the last decade a series of experiments has been performed on elastic
scattering of unstable nuclei at energies around the Coulomb barrier. It has
been found that the scattering cross sections show a significantly different
behavior for weakly bound projectiles compared to tightly bound projectiles
like e.g.\ the \al\ particle. The small binding energy of valence nucleons in
orbitals with small angular momentum leads to wave functions which extend to
very large radii, exceeding by far the usual $A^{1/3}$ radius dependence. 
Due to the corresponding long range absorption the Fresnel diffraction peak in
the elastic scattering angular
distribution is damped, and the elastic scattering cross section at backward
angles is relatively small. As a consequence, the derived total reaction cross
section \str\ for these exotic nuclei (e.g.\ \hes ) is much larger than for 
tightly bound projectile (e.g.\ \al\ particle) induced reactions. Fusion,
breakup, and transfer reactions have been studied as the relevant reaction
mechanisms. 

As one focus on elastic scattering experiments with \hes, results have been
reported for heavy target nuclei like $^{197}$Au, $^{208}$Pb, and $^{209}$Bi
\cite{War95,Kol98,Agu00,Agu01,Kak03,Kak06,San08} and intermediate mass nuclei
like $^{64}$Zn and $^{65}$Cu \cite{Nav04,Pie04,Cha08}. Some data are also
available for lighter target nuclei like $^{12}$C (e.g.,
\cite{milin,Kee08a}). In addition, elastic scattering of $^{11}$Be has been
studied recently \cite{Maz06,Maz07,Pie10}. For a complete list of references,
see the recent reviews \cite{Kee07,Kee09}.

Moreover, a series of theoretical investigations
\cite{Mohr00a,Rus03,Kee03,Abu04,Rus05,Mor07,Bor07,Kee08b,Mac09,Kuc09,Ers10,Fer10,Luk10}
on \hes\ elastic scattering
has been performed in the last years; they are also summarized in the review
articles by Keeley and coworkers \cite{Kee07,Kee09}. The present study
reanalyzes recently published data of the \snt \rhes \snt\ elastic scattering
cross section \cite{Far10} which filled the gap between targets with $A \ll
100$ and $A \approx 200$. We compare these results to \snt \raa \snt\ elastic
scattering data which have been obtained in the same experiment. The
similarities and the differences of the weakly bound projectile \hes\ and the
tightly bound projectile \al\ are nicely visible in this comparison.

The present study uses double folding potentials for the real part of the
potential; this type of potentials is widely used in literature. The imaginary
part is parametrized by Woods-Saxon potentials. The parameters of the
potentials are restricted by the systematics of volume integrals which was
found for many \al -nucleus systems \cite{Atz96}; this systematics was
successfully extended to \hes\ in \cite{Mohr00a,Agu01}. Further information on
the $^{120}$Sn-\al\ potential is obtained from the analysis of angular
distributions at higher energies \cite{Kum68,Bar66,Bar65,Kut01} and excitation
functions at lower energies \cite{Tab75,Bad78}.

The most important quantity for the description of elastic scattering data
below and around the Coulomb barrier are the reflexion coefficients $\eta_L$
which define the total reaction cross section. There is a characteristic
increase of the $\eta_L$ from $\eta_L \approx 0$ (i.e.\ almost complete
absorption) for small angular momenta $L$ to $\eta_L \approx 1$ (i.e.\ no
absorption) for large $L$ corresponding to large impact parameters in a
classical picture. It will be shown that the dependence of $\eta_L$ on the
angular momentum $L$ differs significantly for \snt \rhes \snt\ and \snt \raa
\snt\ elastic scattering. This difference can be considered as a new criterion
for unusual strong absorption because of the halo nature of \hes .

This article is organized as follows: In Sect.~\ref{sec:exp} we repeat very
briefly a discussion of the experimental set-up which is identical to
\cite{Far10}. Sect.~\ref{sec:opt} contains an optical model (OM) analysis of
the \snt \rhes \snt\ (Sect.~\ref{sec:om_he6}) and \snt \raa
\snt\ (Sect.~\ref{sec:om_a}) scattering data and a discussion of the results
(Sect.~\ref{sec:om_disc}). Finally, conclusions are drawn in
Sect.~\ref{sec:sum}. Energies are given in the center-of-mass (c.m.) system
except explicitly noted as laboratory energy $E_{\rm{lab}}$.

\section{Experimental technique}
\label{sec:exp}
The scattering experiment has been performed at the 8UD S{\~a}o Paulo
Pelletron Laboratory at the RIBRAS (Radioactive Ion Beams in Brazil)
facility \cite{ribras}. A primary $^7$Li$^{3+}$ beam with energies around
25\,MeV and a beam 
current of 300\,nAe hits the primary ${^9}$Be target. The reaction products
are collimated and enter a solenoid which focuses the primary $^7$Li particles
onto a ``lollipop'' where the $^7$Li particles
are stopped. Because of the different magnetic rigidity, the secondary
\hes\ and \al\ particles do not hit the ``lollipop'', but reach the secondary
\snt\ target. Typical beam intensities are about $10^4 - 10^5$ particles per
second at the secondary target position. A $3.8$\,mg/cm$^2$ isotopically
enriched ($98.29\,\%$) $^{120}$Sn target 
and a $3.0$\,mg/cm$^2$ $^{197}$Au target have been used as secondary
targets. As the scattering $^4$He+$^{197}$Au 
is pure Rutherford at forward angles in the energies of the present
experiment, runs with gold target have been  
performed just before and after every $^{120}$Sn run in order to normalize the
$^4$He+$^{120}$Sn cross sections \cite{Far10}.

The scattered particles are detected and
identified in a system of $\Delta E$ and $E$ silicon detectors. A schematic
view of the set-up is given in Fig.~1 of \cite{Far10}.

The \hes\ beam is produced by one-proton removal from $^7$Li in the
$^9$Be($^7$Li,\hes )$^{10}$B reaction. But also the reaction $^9$Be($^7$Li,\al
)$^{12}$B may occur in the primary target, leading to an \al\ contamination of
the secondary beam. Because of the much larger $Q$-value of the \al -producing
reaction ($Q_\alpha = +10.5$\,MeV compared to $Q_{^6{\rm{He}}} = -3.4$\,MeV) the
\al\ particles have slightly higher energies around 30\,MeV. The \al\ beam
contamination is clearly visible in the $\Delta E$--$E$ spectra in Fig.~2 of
\cite{Far10}. This contamination can be used to measure the \snt \raa
\snt\ elastic scattering cross section simultaneously with the \snt \rhes
\snt\ experiment.

\begin{figure}
\includegraphics[bbllx=55,bblly=35,bburx=525,bbury=770,width=\columnwidth,clip=]{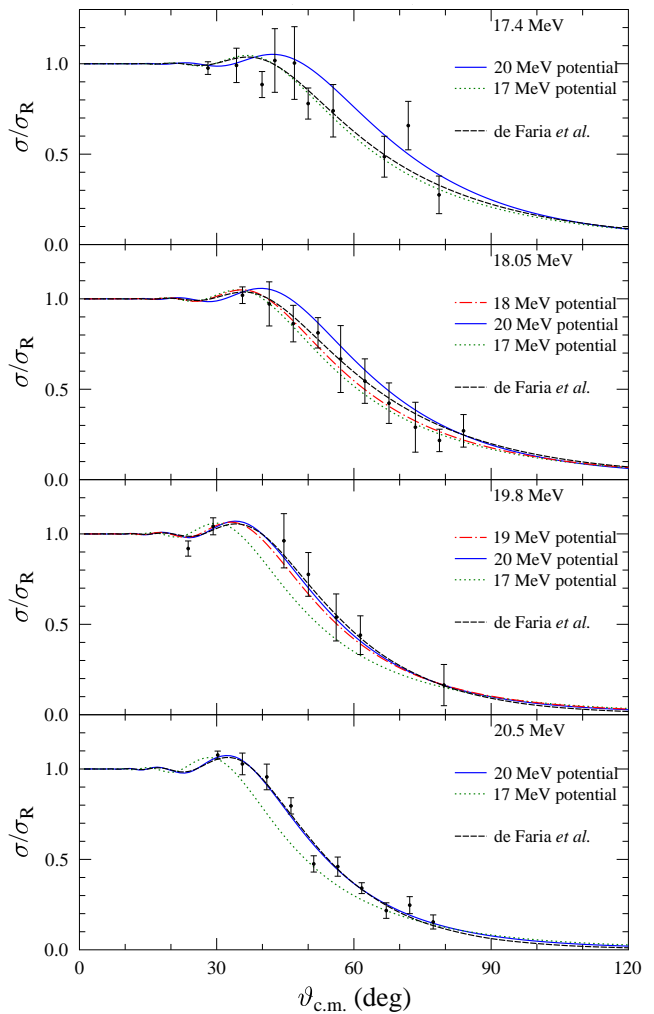}
\caption{
\label{fig:scat_he6}
(Color online)
Rutherford normalized elastic scattering cross sections of
\snt \rhes \snt\ reaction at $E_{\rm{lab}}$ = 17.4, 18.05, 19.8, and 20.5\,MeV
versus the scattering angle $\vartheta_{\rm{c.m.}}$ in the center-of-mass
system (from \cite{Far10}). The black dashed lines are the results from the
original analysis in \cite{Far10}. The blue full lines are obtained from the
fit to the 20\,MeV data, and the green dotted lines are obtained from the fit
to the 17\,MeV data. The dash-dotted red lines are the interpolations for the
18 and 19\,MeV data. The parameters of the fits are listed in Table
\ref{tab:pot_he6}. Further discussion see text (Sect.~\ref{sec:om_he6}).
}
\end{figure}

\begin{figure}
\includegraphics[bbllx=55,bblly=25,bburx=525,bbury=785,width=\columnwidth,clip=]{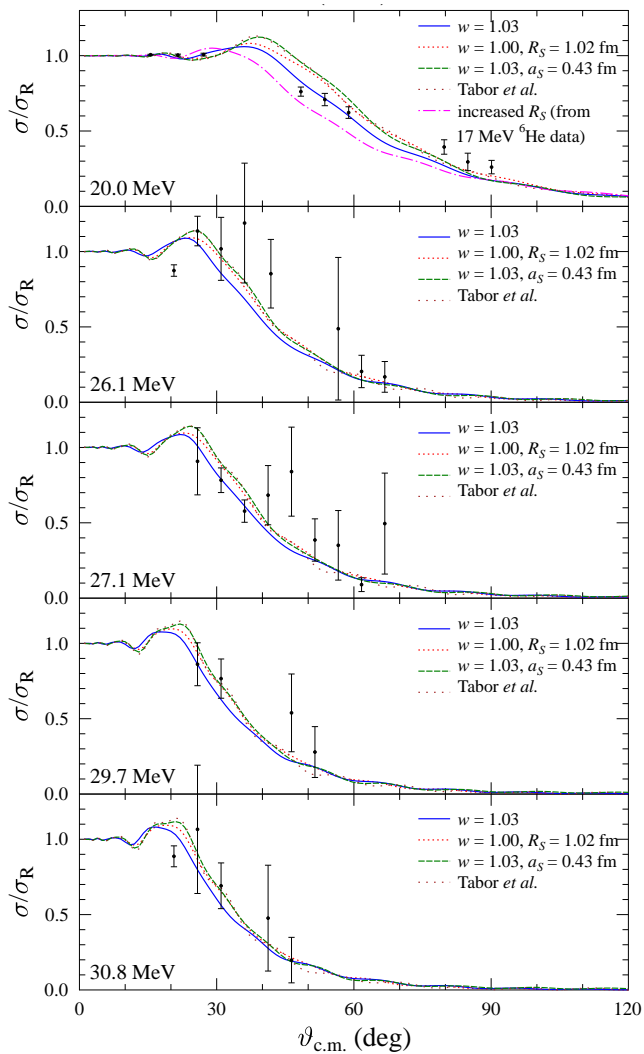}
\caption{
\label{fig:scat_a_low}
(Color online)
Rutherford normalized elastic scattering cross sections of the
\snt \raa \snt\ reaction at $E_{\rm{lab}}$ = 20.0, 26.1, 27.1, 29.7,
and 30.8\,MeV versus the scattering angle $\vartheta_{\rm{c.m.}}$ in
center-of-mass system. The lines are the results from the optical model
calculations in Sect.~\ref{sec:om_a} using different width parameters $w$ of
the real part and different imaginary radii $R_S$ and diffusenesses $a_S$ as
indicated in the 
figure. In addition, the influence of an increased imaginary radius $R_S$ is
shown for the 20\,MeV data. The calculation of Tabor {\it et al.} \cite{Tab75}
was adjusted to reproduce low-energy excitation functions (see
Sec.~\ref{sec:exci}). Further details see text (Sect.~\ref{sec:om_a}). 
}
\end{figure}

The result of the previous \snt \rhes \snt\ experiment \cite{Far10} is shown
in Fig.~\ref{fig:scat_he6} together with the original analysis of \cite{Far10}
and the new analysis which is discussed in the following
Sect.~\ref{sec:om_he6}. The new \snt \raa \snt\ elastic scattering data are
shown in Fig.~\ref{fig:scat_a_low} together with the theoretical results of
this work. Except for the $20$ MeV data which was 
obtained in a previous $^8$Li$+^{120}$Sn experiment \cite{pfariathesis}, the
laboratory energies of the \al-beams
are related to the \hes\ energies by 
$E_{\alpha,{\rm{lab}}}=\frac{3}{2} E_{^6{\rm{He}},{\rm{lab}}}$
due to the band-pass of the solenoid ($ B \rho=\sqrt{2mE_{\rm{lab}}}/q$).

\section{Optical model analysis}
\label{sec:opt}
The complex optical model potential (OMP) is given by:
\begin{equation}
U(r)\,=V_{C}(r)+V(r)+iW(r),
\end{equation}
where \textit{V$_C$(r)} is the Coulomb potential, \textit{$V(r)$}, and
\textit{$W(r)$} are the real and the imaginary parts of the nuclear potential,
respectively. The real part of the potential is calculated from the folding
procedure \cite{Kob84,Sat79} using a density-dependent nucleon-nucleon
interaction. The calculated folding potential is adjusted to the experimental
scattering data by two parameters
\begin{equation}
V(r) = \lambda \, V_{F}(r/w)
\label{eq:fold}
\end{equation}
where $\lambda \approx 1.1 - 1.4$ is the potential strength parameter
\cite{Atz96} and $w \approx 1.0 \pm 0.05$ is the width parameter that
slightly modifies the potential width. (Larger deviations of the width
parameter $w$ from unity would indicate a failure of the folding potential.)
The nuclear densities of \snt\ and \al\ are derived from the measured
charge density distributions which are compiled in \cite{Vri87}: For \snt\ the
three-parameter Gaussian distribution \cite{Fic72} is used. Almost identical
folding potentials are obtained from the second available density distribution
for \snt\ \cite{Bar67} which has been measured earlier in a smaller range of
momentum 
transfers. For the \al\ particle the sum-of-Gaussian parameterization of
\cite{Sick82} is used. The \hes\ density is taken from the $^6$Li density
determined in \cite{Li71}; both nuclei \hes\ and $^6$Li have 2 nucleons in the
$p$-shell with similar separation energies. This density has been
applied successfully in the calculation of $^{209}$Bi\rhes $^{209}$Bi elastic
scattering \cite{Mohr00a,Agu01}. Limitations of this choice may become visible
in the width parameter $w$ of the real part of the potential. However, a very
similar folding potential is obtained from recently published theoretical
densities of \hes\ \cite{Bri10}; the consequences of the different choices for
the \hes\ density will be studied in a subsequent paper. For further details
of the folding potential, see also \cite{Abe93,Mohr97}. 

The imaginary part $W(r)$ is taken in the usual Woods-Saxon
parametrization. For the fits to the experimental data we use 
volume and surface potentials:
\begin{equation}
W(r) = W_V \times f(x_V) + 4 \, W_S \times \frac{df(x_S)}{dx_S}
\end{equation}
with the potential depths $W_V$ and $W_S$ of the volume and surface parts and 
\begin{equation}
f(x_i) = \frac{1}{1+\exp{(x_i)}}
\end{equation}
and $x_i = [r-R_i\,(A_P^{1/3}+A_T^{1/3})]/a_i$ with the radius parameters $R_i$
in the heavy-ion convention, the diffuseness parameters $a_i$, and $i=S,V$. It
is well established that at very low energies the surface contribution of the
imaginary part is dominating; e.g., in \cite{Gal05} it is suggested that the
surface contribution is about 80\,\% for \al\ scattering of the neighboring
nuclei $^{112}$Sn and $^{124}$Sn at energies below 20\,MeV. At higher
energies, i.e.\ significantly above the Coulomb barrier, the volume
contribution is dominating. 

The Coulomb potential $V_C(r)$ is taken in the usual form of a homogeneously
charged sphere. The Coulomb radius $R_C$ is taken from the root-mean-square
(rms) radius of the real folding potential with $w = 1.0$; the sensitivity of
the calculations on minor changes of $R_C$ is negligible.

For a fit to few data points of elastic scttering around the Coulomb barrier,
the number of adjustable parameters should be as small as possible because
there are significant ambiguities for the derived potentials; the underlying
problem is that the elastic scattering cross section is sensitive to the phase
shifts and reflexion coefficients which are properties of the wave function
far outside the nuclear radii: ($i$) the so-called ``family problem'' is a
discrete ambiguity where real potentials with different depths lead to a
similar description of the scattering data because the wave functions are very
similar in the exterior, whereas in the interior the number of nodes may
change. ($ii$) Continuous ambiguities are found: e.g., a larger potential
depth may be compensated by a smaller radius parameter, leading to more or
less the same total potential strength and thus to the same wave function in
the exterior region. In some cases this leads to a so-called ``one-point
potential'' (e.g.\ \cite{Bad78,Mohr97,Sig00,Fer10}).

For a reduction of the adjustable parameters we use the systematic behavior of
the volume integrals of the potentials which has been found in
\cite{Atz96,Mohr00a}. For intermediate mass and heavy nuclei the volume
integrals $J_R$ of the real part of the potential are practically independent
of the chosen nuclei and depend only weakly on energy with a
maximum around 30\,MeV. A Gaussian parameterization has been suggested in
\cite{Mohr00b} for energies below and slightly above the maximum of $J_R$ at
$E_{R,0} = 30$\,MeV:
\begin{equation}
J_R(E) = J_{R,0} \times \exp{\Big[-{\frac{(E-E_{R,0})^2}{\Delta_R^2}}\Big]}
\label{eq:jr}
\end{equation}
with the maximum value $J_{R,0} = 350$\,MeV\,fm$^3$ and the width $\Delta_R =
75$\,MeV. Potentials with $J_R$ from Eq.~(\ref{eq:jr}) have been used for
\al\ scattering \cite{Atz96}, \al\ decay \cite{Mohr00b}, and \hes\ scattering
\cite{Mohr00a,Agu01}. The energy dependence of $J_R$ is weak; e.g.,
$J_R$ changes by only a few per cent in the considered energy range of this
work. (Note that the negative signs of the volume integrals are, as usual,
neglected in the discussion.)

Contrary to the real volume integrals $J_R$, the imaginary volume integrals
$J_I$ depend on the chosen nuclei and on energy. The energy dependence of
$J_I$ has been parametrized according to Brown and Rho \cite{Bro81}
\begin{equation}
J_I(E) = J_{I,0} \times \frac{(E-E_{I,0})^2}{(E-E_{I,0})^2 + \Delta_I^2}
\label{eq:ji}
\end{equation}
with a saturation value $J_{I,0}$, the threshold value $E_{I,0} = 1.171$\,MeV
(corresponding to the first excited $2^+$ state in \snt ), and the slope
parameter $\Delta_I$. Saturation values around $J_{I,0} \approx
100$\,MeV\,fm$^3$ have been found in \al\ scattering with a trend to smaller
$J_{I,0}$ for doubly-magic targets and increasing $J_{I,0}$ for semi-magic or
non-magic targets. For the combination of a semi-magic \hes\ projectile and a
semi-magic $^{209}$Bi target $J_{I,0} = 127$\,MeV\,fm$^3$ and $\Delta_I =
12.7$\,MeV were found \cite{Mohr00a}; these values are adopted for the
analysis of \snt \rhes \snt\ elastic scattering which is also a combination of
a semi-magic projectile and a semi-magic target. For \snt \raa \snt\ elastic
scattering a smaller saturation value of $J_{I,0} = 80$\,MeV\,fm$^3$ is used
which is derived from scattering data at higher energies (see
Sect.~\ref{sec:om_a}).

From the above considerations the volume integrals $J_R$ and $J_I$ for the
analysis of \snt \rhes \snt\ and \snt \raa \snt\ elastic scattering are
fixed. Hence the two adjustable parameters in the real part (strength
parameter $\lambda$ and width parameter $w$) are related by the volume
integral $J_R$ in Eq.~(\ref{eq:jr}), and the three adjustable Woods-Saxon
parameters (depth $W_V$ or $W_S$, radius $R_V$ or $R_S$, and diffuseness $a_V$
or $a_S$) are related by the volume integral $J_I$ in Eq.~(\ref{eq:ji}).

\subsection{\snt \rhes \snt }
\label{sec:om_he6}
In addition to the above restrictions for the volume integrals $J_R$ and
$J_I$, we fix the imaginary surface diffuseness to a standard value $a_S =
0.7$\,fm. The small volume part of the imaginary potential at low energies
\cite{Gal05} is neglected: $W_V = 0$.

In a next step we adjust the remaining parameters to the \snt \rhes
\snt\ scattering data at $E_{\rm{lab}} = 20.5$\,MeV (referred to as ``20\,MeV
data'' in the following; the same convention of referring to the integer part
of the laboratory energy $E_{\rm{lab}}$ will be used for all data). An
excellent description of the 20\,MeV data is found (see
Fig.~\ref{fig:scat_he6}, full blue line) using a relatively small width
parameter of $w =0.95$ (see also Sect.~\ref{sec:om_disc}).
The same potential is now applied to the measured angular distributions at
lower energies. Increasing discrepancies are observed for lower energies
(Fig.~\ref{fig:scat_he6}, full blue lines): the calculated cross section at
backward angles is larger than the measured values. 

Because of the minor energy dependence of the real potential, the width
parameter $w$ was fixed now, and we tried to fit the lowest 17\,MeV data by a
readjustment of the imaginary part of the potential with a fixed $J_I$ from
Eq.~(\ref{eq:ji}). A clear increase of the radius parameter $R_S$ by about
15\,\% was found; then an excellent description of the 17\,MeV data can be
obtained. This 17\,MeV potential is not able to describe the
angular distributions at the other energies, where the calculated cross sections
underestimate the experimental results at backward angles
(Fig.~\ref{fig:scat_he6}, dotted green lines).

Finally, we interpolate the imaginary radius parameter $R_S$ between the
17\,MeV and the 20\,MeV results and use it for the remaining 18\,MeV and
19\,MeV angular distributions. An excellent agreement is obtained for all
measured angular distributions (Fig.~\ref{fig:scat_he6}, dash-dotted red
lines). The resulting parameters of the potentials are listed in Table
\ref{tab:pot_he6}.
\begin{table*}
  \caption{\label{tab:pot_he6}
    Parameters of the potentials of \snt \rhes \snt\ elastic scattering in
    Fig.~\ref{fig:scat_he6}.
}
\begin{center}
\begin{tabular}{ccccccccccc}
\multicolumn{1}{c}{$E_{\rm{lab}}$} 
& \multicolumn{1}{c}{$\lambda$} 
& \multicolumn{1}{c}{$w$\footnote{fixed value, adjusted to the 20\,MeV data}} 
& \multicolumn{1}{c}{$J_R$\footnote{from Gaussian parameterization,
    Eq.~(\ref{eq:jr})}} 
& \multicolumn{1}{c}{$r_{R,rms}$} 
& \multicolumn{1}{c}{$J_I$\footnote{from Brown-Rho parameterization,
    Eq.~(\ref{eq:ji})}}
& \multicolumn{1}{c}{$r_{I,rms}$} 
& \multicolumn{1}{c}{$W_{S}$}
& \multicolumn{1}{c}{$R_{S}$}
& \multicolumn{1}{c}{$a_S$\footnote{fixed value}}
& \multicolumn{1}{c}{\str } \\
\multicolumn{1}{c}{(MeV)} 
& & 
& \multicolumn{1}{c}{(MeV\,fm$^3$)} 
& \multicolumn{1}{c}{(fm)} 
& \multicolumn{1}{c}{(MeV\,fm$^3$)} 
& \multicolumn{1}{c}{(fm)} 
& \multicolumn{1}{c}{(MeV)}
& \multicolumn{1}{c}{(fm)} 
& \multicolumn{1}{c}{(fm)}
& \multicolumn{1}{c}{(mb)} \\
\hline
17.40  & 1.207  & 0.95 & 339.0 & 5.477
& 75.6  & 9.320  & 19.2  & 1.315  & 0.7 & 1479 \\
18.05  & 1.210  & 0.95 & 339.9 & 5.477
& 78.0  & 9.074  & 21.0  & 1.277  & 0.7 & 1503 \\
19.80  & 1.219  & 0.95 & 342.4 & 5.477
& 83.8  & 8.415  & 26.6  & 1.174  & 0.7 & 1538 \\
20.50  & 1.222  & 0.95 & 343.2 & 5.477
& 85.9  & 8.153  & 29.3  & 1.133  & 0.7 & 1546 \\
\hline
\end{tabular}
\end{center}
\end{table*}

The total reaction cross sections \str\ can be calculated from the
reflexion coefficients $\eta_L$. We find that \str\ decreases only
slightly with energy from \str\ = 1546\,mb at the highest energy
$E_{\rm{lab}} = 20.5$\,MeV to \str\ = 1479\,mb at the lowest energy of
$E_{\rm{lab}} = 17.4$\,MeV (see Table \ref{tab:pot_he6}). These results agree
with the original optical model analysis of \cite{Far10} within less than
5\,\%. 

For comparison, Fig.~\ref{fig:scat_he6} shows also the original analysis of
\cite{Far10} using Woods-Saxon potentials without any restriction (black
dashed lines). It is obvious that the systematic potentials from this work are
able to reproduce the measured angular distributions with the same quality as
the unrestricted Woods-Saxon potentials which do not show any systematic
bahavior; their volume integrals $J_R$ and $J_I$ vary strongly with energy.

\subsection{\snt \raa \snt }
\label{sec:om_a}
The analysis of  the \snt \raa \snt\ system elastic scattering benefits from
the fact that three angular distributions have been measured at higher
energies \cite{Kum68,Bar66,Bar65,Kut01}. These
angular distributions can be used to fix the real part of the optical
potential with small uncertainties. Thus, the number of adjustable parameters
in the analysis of the new angular distributions at lower energies (see
Fig.~\ref{fig:scat_a_low}) is reduced, and the imaginary part can be deduced
from the experimental data for a subsequent comparison with the
\hes\ case. Further information on the potential can be obtained from the
analysis of excitation functions which have been measured at lower energies
\cite{Tab75,Bad78}. 

Data at higher energies can be best reproduced using an imaginary potential of
Woods-Saxon volume type. Somewhat arbitrary, we take the three data sets from
literature at $E_{\rm{lab}} = 34.4$\,MeV \cite{Kum68}, 40.0\,MeV
\cite{Bar66,Bar65}, 
and 50.5\,MeV \cite{Kut01} as the ``high-energy'' data which are analyzed with
a volume Woods-Saxon imaginary part, whereas our new data below 30\,MeV are
analyzed as ``low-energy'' data using a surface Woods-Saxon imaginary
part. Obviously, there must be an intermediate energy range with the
transition from surface Woods-Saxon to volume Woods-Saxon potentials. This
transitional region is around the 34\,MeV data of \cite{Kum68}; however, these
data are not adequate for a precise determination of the optical potential
(see below).

\subsubsection{Angular distributions above approx.\ 30\,MeV}
\label{sec:a_high}
Three angular distributions of \snt \raa \snt\ elastic scattering have been
published. The data by Kuterbekov {\it et al.}\ \cite{Kut01} have been
measured at $E_{\rm{lab}} = 50.5$\,MeV. The data cover an angular range from
about $10^\circ$ to $60^\circ$. The numerical data are available in the EXFOR
data base, but no further information on the experiment is available. The data
of Baron {\it et al.}\ \cite{Bar66} are described in detail in an earlier
report \cite{Bar65}, including the numerical data with statistical
errors. Because of very tiny statistical error bars in the forward direction
of far below 1\,\%, a systematic error of 5\,\% has been added quadratically
to all data points. In addition, the given energy of $E_\alpha = 40.00 \pm
0.25$\,MeV \cite{Bar65} has been reduced to an effective energy $E_{\rm{lab}}
= 39.95$\,MeV because of the energy loss in the target. This angular
distribution covers almost the full angular range from about $20^\circ$ to
$150^\circ$. Finally, the data of Kumabe {\it et al.}\ \cite{Kum68} cover only
a very limited angular range from about $20^\circ$ to $60^\circ$. The data
have been extracted from Fig.~2 of \cite{Kum68} which shows the absolute cross
sections without error bars. Because of the limited angular range, the
uncertainties of the digitization procedure, and the missing error bars, any
fit of these data has significant uncertainties.

The three angular distributions have been fitted using two adjustable
parameters in the real part (strength parameter $\lambda$ and width parameter
$w$) and three parameters in the imaginary part (depth $W_V$, radius $R_V$,
and diffuseness $a_V$). Additionally, the absolute values of the measured
cross sections were allowed to vary. It is well-known that the cross sections
at forward directions do practically not depend on the underlying potentials;
in particular, at very forward directions the cross section approaches the
Rutherford cross section for all optical potentials. Thus, it is common
practice to normalize the measured data to calculated values at forward
directions because an absolute measurement requires the absolute determination
of the target thickness and uniformity, detector solid angle, and beam current
and a proper deadtime correction. The scaling factor $s$ for the correction of
the experimental data is defined by $\sigma_{\rm{exp}}^{\rm{corr}} = s \times
\sigma_{\rm{exp}}^{\rm{raw}}$ where $\sigma_{\rm{exp}}^{\rm{raw}}$ are the
published cross section data.  It has been stated e.g.\ in \cite{Kum68} that
this theoretical normalization $s$ deviates by $10 - 25$\,\% from unity for
the tin targets used in that experiment. It is interesting to note that the
obtained potential parameters are not very sensitive to the scaling factor $s$
as long as $s$ remains far below a factor of two because the diffraction
pattern in the experimental data at higher energies nicely defines the
underlying potential.

\begin{figure}
\includegraphics[bbllx=35,bblly=10,bburx=530,bbury=755,width=\columnwidth,clip=]{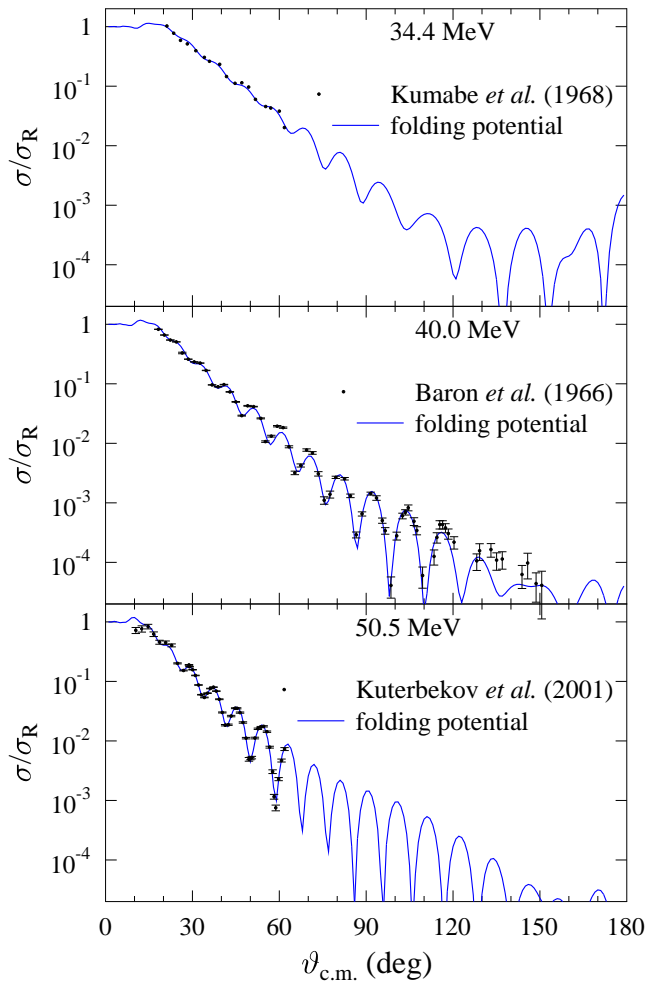}
\caption{
\label{fig:scat_a_high}
(Color online)
Rutherford normalized elastic scattering cross sections of
\snt \raa \snt\ reaction at higher energies $E_{\rm{lab}}$ = 34.4, 40.0, and
50.5\,MeV \cite{Kum68,Bar66,Bar65,Kut01} versus the scattering angle
$\vartheta_{\rm{c.m.}}$ in center-of-mass frame. The calculated angular
distributions use a double-folding potential in the real part and a volume
Woods-Saxon potential in the imaginary part. Further details see text.
}
\end{figure}

The results of the analysis are shown in Fig.~\ref{fig:scat_a_high}, and the
obtained parameters are listed in Table \ref{tab:pot_a_high}. An excellent
agreement between the scaled experimental data and the theoretical analysis is
found for all energies under study.

\begin{table*}
  \caption{\label{tab:pot_a_high}
    Parameters of the potentials of \snt \raa \snt\ elastic scattering at
    higher energies above 30\,MeV in Fig.~\ref{fig:scat_a_high}.
}
\begin{center}
\begin{tabular}{cccccccccccc}
\multicolumn{1}{c}{$E_{\rm{lab}}$} 
& \multicolumn{1}{c}{$s$} 
& \multicolumn{1}{c}{$\lambda$} 
& \multicolumn{1}{c}{$w$} 
& \multicolumn{1}{c}{$J_R$}
& \multicolumn{1}{c}{$r_{R,rms}$} 
& \multicolumn{1}{c}{$J_I$}
& \multicolumn{1}{c}{$r_{I,rms}$} 
& \multicolumn{1}{c}{$W_{V}$}
& \multicolumn{1}{c}{$R_{V}$}
& \multicolumn{1}{c}{$a_V$}
& \multicolumn{1}{c}{$\sigma_{\rm{reac}}$} \\
\multicolumn{1}{c}{(MeV)} 
& & & 
& \multicolumn{1}{c}{(MeV\,fm$^3$)} 
& \multicolumn{1}{c}{(fm)} 
& \multicolumn{1}{c}{(MeV\,fm$^3$)} 
& \multicolumn{1}{c}{(fm)} 
& \multicolumn{1}{c}{(MeV)}
& \multicolumn{1}{c}{(fm)} 
& \multicolumn{1}{c}{(fm)} 
& \multicolumn{1}{c}{(mb)} \\
\hline
34.4  & 0.97  & 1.222  & 1.029 & 340.5 & 5.461
& 58.6  & 6.496  & 12.9  & 1.213  & 0.583  & 1742 \\
40.0  & 1.48  & 1.190  & 1.048 & 347.4 & 5.561
& 77.2  & 6.495  & 16.7  & 1.173  & 0.543  & 1927 \\
50.5  & 1.18  & 1.227  & 1.018 & 324.1 & 5.407
& 78.6  & 6.315  & 18.2  & 1.197  & 0.490  & 1939 \\
\hline
\end{tabular}
\end{center}
\end{table*}

After a minor scaling of less than 20\,\% ($s = 1.18$) the 50\,MeV data can be
described very well except the two data points at most forward angles. Because
of the reproduction of the diffraction pattern over the full measured angular
range, it seems to be very unlikely that there is such a huge deviation
between theory and experiment at small angles around $15^\circ$. It should be
kept in mind that the error bars in \cite{Kut01} are statistical only; however,
because of the strong angular dependence of the Rutherford cross section, the
uncertainties of data points at forward angles are usually defined by
systematic uncertainties (e.g.\ from the angular calibration or the deadtime
correction).

The data of Baron {\it et al.}~\cite{Bar66,Bar65} cover almost the full angular
range and are thus an ideal data set for the determination of the optical
potential. The reproduction of the angular distribution is excellent over the
full angular range. However, a significant scaling of the data ($s = 1.48$)
was necessary; this seems to be justified because otherwise the most forward
data point at $18^\circ$ deviates by almost a factor of two from the
Rutherford cross section.

As pointed out above, the data at 34\,MeV \cite{Kum68} have less explanatory
power. Here a small scaling factor of $s = 0.97$ is found. The reason for this
$s \approx 1$ is simply that \cite{Kum68} have already applied the same
normalization procedure to their data. The found deviation of 3\,\% thus
simply provides an estimate for the uncertainty of the digitization procedure
which had to be used to extract the data from their Fig.~2.

From the obtained parameters (see Table \ref{tab:pot_a_high}) the following
conclusions can be drawn. The real part of the potential behaves very
regularly with the expected decrease of the real volume integral $J_R$ at
higher energies \cite{Atz96}. The resulting $J_R$ remain close to the suggested
Gaussian parameterization in Eq.~(\ref{eq:jr}) although this parameterization
is not expected to remain valid far above the maximum around 30\,MeV
\cite{Mohr00b}. The width parameter $w$ is always slightly above
1.0; thus, for the following calculations at lower energies we adopt the
average value of $\bar{w} = 1.032$. Together with the parameterization of
$J_R$ at low energies in Eq.~(\ref{eq:jr}), the real part of the optical
potential is completely fixed now. The imaginary part increases with energy
and saturates at $J_{I,0} \approx 80$\,MeV\,fm$^3$. As expected, this value is
somewhat smaller than the result for $^6$He ($J_{I,0} = 127$\,MeV\,fm$^3$). The
available data are not sufficient to derive the slope parameter $\Delta_I$ of
the Brown-Rho parameterization in Eq.~(\ref{eq:ji}). Instead, we use the same
value $\Delta_I = 12.7$\,MeV for \al\ and \hes\ in this paper.

The relatively large value of $w = 1.032$ from the \snt \raa \snt\ data at
higher energies together with the small value of $w \approx 0.95$ derived from
the \snt \rhes \snt\ data indicates that there is no major problem with the
underlying \snt\ density which should show up as modification for $w$ in the
same direction in both experiments. This is not surprising because the
\snt\ charge density has been measured in two independent experiments
\cite{Fic72,Bar67}, and there is no evidence for a peculiar behavior of the
neutron density (e.g.\ neutron skin) in \snt\ \cite{War10,Oze09}. Instead, it
may be concluded that the chosen \hes\ density is not very
precise. Surprisingly, this problem was not found in the analysis of
$^{209}$Bi\rhes $^{209}$Bi scattering data \cite{Mohr00a,Agu01}; however, it
may have been masked there by the larger Coulomb barrier of $^{209}$Bi.

The largest width parameter $w = 1.048$ was obtained from the analysis of the
40\,MeV angular distribution of \cite{Bar65,Bar66}. A smaller width parameter
of $w \approx 1.02$, closer to unity and in better agreement with the other
data, can be obtained if the energy is changed to 42\,MeV instead of
40\,MeV. It is interesting to note that the authors of \cite{Bar65,Bar66}
later refer to their data as ``42-MeV scattering data'' \cite{Bar71} whereas
in \cite{Bar65} it is explicitly stated that ``the incident beam energy is
$40.00 \pm 0.25$\,MeV''.

\subsubsection{Angular distributions below approx.\ 30\,MeV}
\label{sec:a_low}
After fixing the complete real potential and the imaginary volume integral
$J_I$ as described in the previous section, now we fixed the geometry of the
imaginary part for the low-energy data below $\approx 30$\,MeV. Because of the
dominating volume term at higher energies and the dominating surface term at
lower energies (e.g.~\cite{Gal05}), it is impossible to use at low energies the 
same geometry of the imaginary potential obtained at higher energies. 
Instead, we follow a procedure
similar to the low-energy \hes\ data. We fix the imaginary surface diffuseness
at $a_S = 0.7$\,fm, and we take the radius parameter $R_S$ from the highest
energy of the \snt \rhes \snt\ data: $R_S = 1.133$\,fm. The depth of the
potential $W_S$ is adjusted to reproduce the volume integral $J_I$ from
Eq.~(\ref{eq:ji}) with the parameters $J_{I,0} = 80$\,MeV\,fm$^3$ and
$\Delta_I = 12.7$\,MeV (as discussed in the previous subsection). As a
consequence, all parameters of the potential are fixed, either to systematics
or to the experimental data at higher energies. The reproduction of the \snt
\raa \snt\ elastic scattering cross section is good for all energies, see
Fig.~\ref{fig:scat_a_low}. The parameters are listed in Table
\ref{tab:pot_a_low}. The total reaction cross section \str\ shows the usual
energy dependence, i.e.\ it increases strongly with increasing energy.
\begin{table*}
  \caption{\label{tab:pot_a_low}
    Parameters of the potentials of \snt \raa \snt\ elastic scattering in
    Figs.~\ref{fig:scat_a_low} and \ref{fig:scat_a_exci}.
}
\begin{center}
\begin{tabular}{rccccccrccr}
\multicolumn{1}{c}{$E_{\rm{lab}}$} 
& \multicolumn{1}{c}{$\lambda$} 
& \multicolumn{1}{c}{$w$\footnote[1]{fixed value from average of high-energy
    data}}
& \multicolumn{1}{c}{$J_R$\footnote[2]{from Gaussian parameterization,
    Eq.~(\ref{eq:jr})}} 
& \multicolumn{1}{c}{$r_{R,rms}$} 
& \multicolumn{1}{c}{$J_I$\footnote[3]{from Brown-Rho parameterization,
    Eq.~(\ref{eq:ji})}}
& \multicolumn{1}{c}{$r_{I,rms}$} 
& \multicolumn{1}{c}{$W_{S}$}
& \multicolumn{1}{c}{$R_{S}$\footnote[4]{fixed value from 20\,MeV \hes\ data}}
& \multicolumn{1}{c}{$a_S$\footnote[5]{fixed value}}
& \multicolumn{1}{c}{\str } \\
\multicolumn{1}{c}{(MeV)} 
& & 
& \multicolumn{1}{c}{(MeV\,fm$^3$)} 
& \multicolumn{1}{c}{(fm)} 
& \multicolumn{1}{c}{(MeV\,fm$^3$)} 
& \multicolumn{1}{c}{(fm)} 
& \multicolumn{1}{c}{(MeV)}
& \multicolumn{1}{c}{(fm)} 
& \multicolumn{1}{c}{(fm)}
& \multicolumn{1}{c}{(mb)} \\
\hline
20.0  & 1.207  & 1.032 & 343.0 & 5.474
& 53.8  & 7.910  & 13.1  & 1.133  & 0.70 & 1121 \\
26.1  & 1.226  & 1.032 & 348.6 & 5.474
& 62.6  & 7.910  & 15.2  & 1.133  & 0.70 & 1663 \\
27.1  & 1.228  & 1.032 & 349.1 & 5.474
& 63.6  & 7.910  & 15.5  & 1.133  & 0.70 & 1727 \\
29.7  & 1.231  & 1.032 & 349.9 & 5.474
& 66.0  & 7.910  & 16.0  & 1.133  & 0.70 & 1870 \\
30.8  & 1.231  & 1.032 & 350.0 & 5.474
& 66.9  & 7.910  & 16.2  & 1.133  & 0.70 & 1923 \\
\hline
$\approx 13.5$\footnote[6]{average energy of excitation functions \cite{Tab75}} 
     & 1.170   & 1.032 & 332.6 & 5.474
& 37.4 & 7.910 & 9.1 & 1.133 
& 0.70
& 150 \\
$\approx 13.5$\footnotemark[6] 
     & 1.170   & 1.032 & 332.6 & 5.474
& 37.4 & 7.588 & 15.1 & 1.133 
& 0.43\footnote[7]{adjusted to excitation functions \cite{Tab75}} & 86 \\
$\approx 13.5$\footnotemark[6] 
     & 1.284   & 1.000\footnotemark[7] & 332.6 & 5.306
& 37.4 & 7.236 & 11.1 & 1.021\footnotemark[7] 
& 0.70 & 61 \\
\hline
\end{tabular}
\end{center}
\end{table*}

In addition, we have studied the sensitivity of the data to minor variations
of the potential. First, a width parameter $w = 1.0$ of the real potential was
used instead of $w = 1.03$ together with a reduced imaginary radius parameter
$R_S = 1.021$\,fm (red dotted line in Fig.~\ref{fig:scat_a_low}; adjusted to
reproduce the excitation functions of \cite{Tab75}, see Sec.~\ref{sec:exci}).
Second, the diffuseness $a_S$ of the imaginary part was decreased to $a_S =
0.43$\,fm instead of 0.7\,fm (green dashed line, again adjusted to reproduce
the excitations of \cite{Tab75}). In both cases the influence on the
scattering cross sections remains relatively small although the 20\,MeV data
around $50^\circ$ are clearly overestimated using $w = 1.0$ and $R_S =
1.021$\,fm or $a_S = 0.43$\,fm from the analysis of the excitation functions.

A significant reduction of the calculated scattering cross section is
found if the increased radius parameter $R_S$ from the 17\,MeV \hes\ data is
taken at the lowest energy of the \al\ data (magenta dash-dotted line). Here
it becomes obvious that the new experimental \snt \raa \snt\ data are not
compatible with the strong increase of the radius parameter $R_S$ at low
energies which was essential for the reproduction of the \snt \rhes
\snt\ data.

\subsubsection{Excitation functions at low energies}
\label{sec:exci}
\begin{figure}
\includegraphics[bbllx=55,bblly=20,bburx=370,bbury=400,width=\columnwidth,clip=]{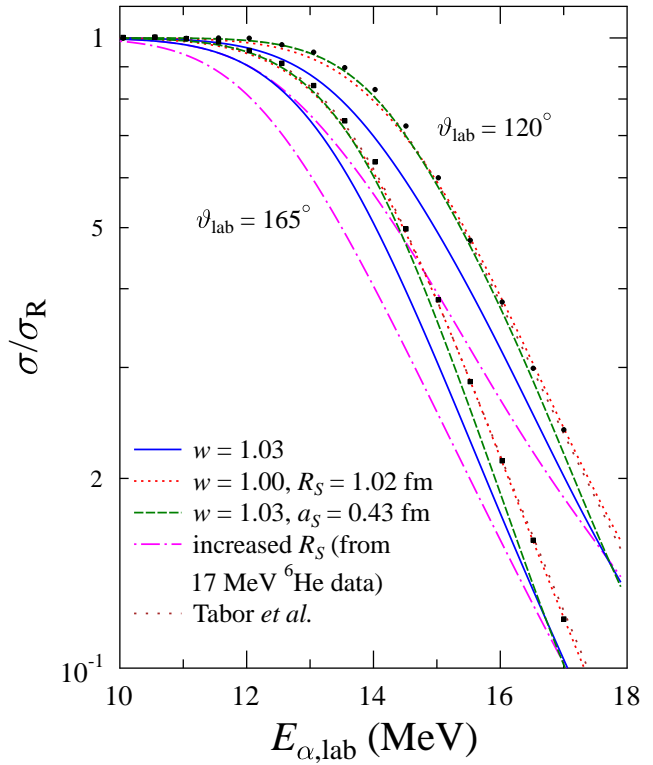}
\caption{
\label{fig:scat_a_exci}
(Color online)
Rutherford normalized excitation function of \snt \raa \snt\ elastic
scattering at $\vartheta_{\rm{lab}} = 120^\circ$ and $165^\circ$ \cite{Tab75}. 
An excellent description of the data at very low energies can be obtained
using either a decreased imaginary diffuseness $a_S = 0.43$\,fm (green dashed)
or a width parameter $w = 1.0$ and a reduced imaginary radius parameter $R_S$
(red dotted), whereas the standard potential slightly underestimates the
measured data (full blue line). For comparison, the original analysis of Tabor
{\it{et al.}}\ \cite{Tab75} is also shown (brown short-dashed, almost
identical to the red dotted line). The increased imaginary radius from the
low-energy \hes\ data is clearly excluded (magenta dash-dotted). Further
discussion see text.
}
\end{figure}

Excitation functions have been measured by Tabor {\it{et al.}}\ and Badawy
{\it{et al.}} \cite{Tab75,Bad78}. Unfortunately, the latter paper only
mentions the measurement and derives a so-called one-point potential, but does
not show the data for \snt \raa \snt ; thus, these data \cite{Bad78} are not
accessible and cannot be used in the analysis. Tabor {\it{et
    al.}}\ \cite{Tab75} show two excitation functions at $\vartheta_{\rm{lab}}
= 120^\circ$ and $165^\circ$ in the energy range from 10 to 17\,MeV in their
Fig.~1. These data are shown together with the original analysis using
a Woods-Saxon potential and the new reanalysis in Fig.~\ref{fig:scat_a_exci}.

In general,
it is not possible to extract an optical potential from low-energy excitation
functions because of ambiguities in the derived potentials. This has been
clearly shown by Badawy {\it et al.}\ \cite{Bad78} in their analysis: ``The
only statement that can be made on the three parameters characterizing a
Woods-Saxon real potential is that they are linked by the relation'' that 
any potential with a depth of 0.2\,MeV at $r = 10.63$\,fm describes
their experimental data. The imaginary potential also cannot be well
determined: ``... the results are very insensitive to the value of $W$
...''. Further details on the one-point potential and its relation to the
so-called ``family problem'' of \al -nucleus potentials are discussed in
\cite{Mohr97} using the precisely determined angular distribution of
$^{144}$Sm\raa $^{144}$Sm at $E \approx 20$\,MeV (see Figs.~5 and 6 of
\cite{Mohr97}). 

Although it is not possible to extract the optical potential, it is
nevertheless possible to test the systematic potentials of this work using the
measured excitation functions of \cite{Tab75}. It is found that the standard
potential with $w = 1.03$, $R_S = 1.133$\,fm, and $a_S = 0.7$\,fm does not
describe the excitation functions at low energies (full blue line in
Fig.~\ref{fig:scat_a_exci}) and underestimates the measured cross
sections. Instead of $a_S = 0.7$\,fm, the diffuseness parameter of the surface 
imaginary part has to be decreased to $a_S = 0.43$\,fm to find reasonable
agreement with the measured excitation functions (green dashed line in
Fig.~\ref{fig:scat_a_exci}). Alternatively, an
excellent description of the data is also obtained using a reduced imaginary 
radius parameter $R_S = 1.021$\,fm, $a_S = 0.7$\,fm, and a width parameter $w
= 1.0$ of the real part; however, such a width parameter $w$ has been excluded
by the high-energy 
angular distributions. This latter result is almost identical to the original
analysis of Tabor {\it{et al.}}\ \cite{Tab75}; similar to that result, the
angular distribution at 20\,MeV is clearly overestimated around $50^\circ$ (see
Fig.~\ref{fig:scat_a_low}). 

Similar to the angular distributions shown in Fig.~\ref{fig:scat_a_low}, a
huge deviation from the measured excitation functions is found if the
increased radius parameter $R_S = 1.315$\,fm is used which has been obtained
from the lowest energy in \snt \rhes \snt\ scattering (dash-dotted magenta
lines in Figs.~\ref{fig:scat_a_low} and \ref{fig:scat_a_exci}).  

The calculated excitation functions may also change when the energy dependence
of the volume integrals $J_R$ and $J_I$ in Eqs.~(\ref{eq:jr}) and
(\ref{eq:ji}) is varied. However, a variation of the Brown-Rho parameters of
the order of 10\,\% has only minor influence on the calculated excitation
functions as long as the geometry of the imaginary potential is not changed.

The parameters of the potentials are also listed in Table \ref{tab:pot_a_low}
at the average energy $E_{\rm{lab}} \approx 13.5$\,MeV of the measured
excitation functions \cite{Tab75}. At this energy both calculations with the
slightly modified standard potential agree
nicely with the measured data (see Fig.~\ref{fig:scat_a_exci}). However, the
preferred calculation with $w = 1.03$ leads to slightly smaller elastic
scattering cross sections which have significant impact on the total reaction
cross section \str : $w = 1.03$ and $a_S = 0.43$\,fm corresponds to \str\ =
86\,mb, $w = 1.0$ and $R_S = 1.021$\,fm corresponds to \str\ = 61\,mb. The
standard potential underestimates the elastic scattering cross sections of
\cite{Tab75} and thus leads to a very high \str\ = 150\,mb. This discrepancy
for \str\ will affect the prediction of \al -induced cross sections in the
statistical model.

\subsection{Discussion}
\label{sec:om_disc}
\begin{figure}
\includegraphics[bbllx=55,bblly=40,bburx=525,bbury=435,width=\columnwidth,clip=]{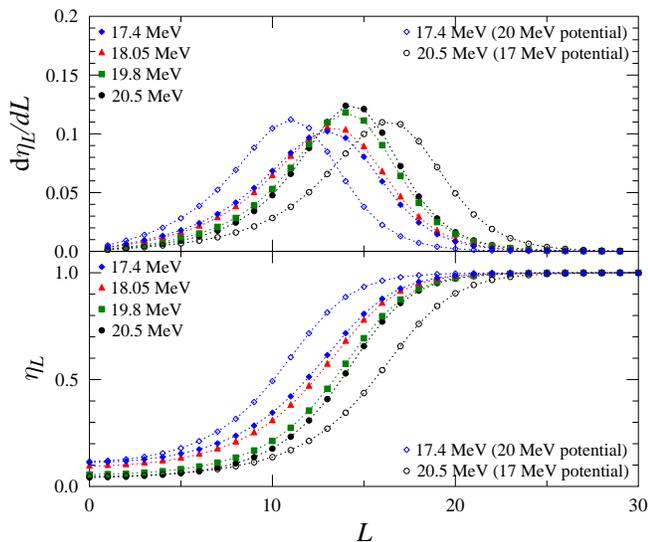}
\caption{
\label{fig:eta_he6}
(Color online)
Reflexion coefficients $\eta_L$ for \snt \rhes \snt\ elastic scattering at
$E_{\rm{lab}}$ = 17.4, 
18.05, 19.8, and 20.5\,MeV (lower part) and the derivatives
$d\eta_L/dL$ (upper part). The full symbols correspond to the
calculations in Fig.~\ref{fig:scat_he6} and Tab.~\ref{tab:pot_he6}; the open
symbols are obtained using the 17\,MeV potential at 20\,MeV and vice
versa. A clear broadening of the derivative $d\eta_L/dL$ at low
energies can be seen. The data points for each $L$ are connected by dotted
lines to guide the eye. Further discussion see text.
}
\end{figure}

\begin{figure}
\includegraphics[bbllx=55,bblly=40,bburx=525,bbury=435,width=\columnwidth,clip=]{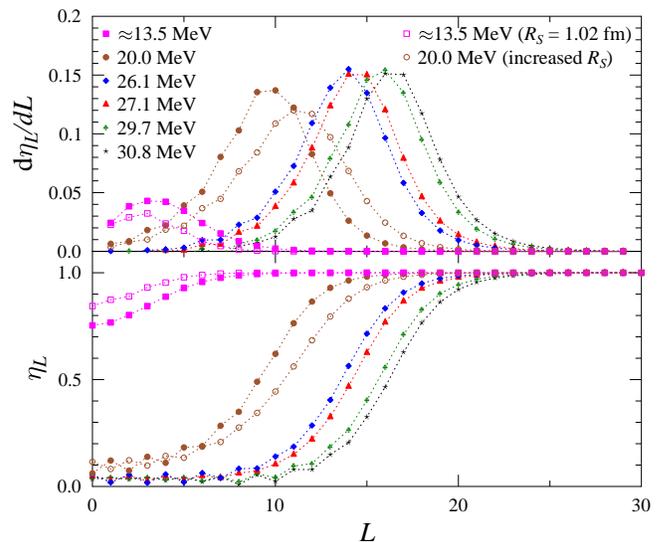}
\caption{
\label{fig:eta_a}
(Color online)
Reflexion coefficients $\eta_L$ for \snt \raa \snt\ elastic scattering at
$E_{\rm{lab}}$ = 20.0, 26.1, 27.1, 29.7, and 30.8\,MeV (lower part) and the
derivatives $d\eta_L/dL$ (upper part). The full symbols correspond to
the calculations in Fig.~\ref{fig:scat_a_low} and Tab.~\ref{tab:pot_a_low}; the
open symbols are obtained at the lowest energy of 20.0\,MeV using an
increased radius $R_S$ of the imaginary surface potential (derived from $^6$He
scattering at the lowest energy). 
Additionally, the results from the excitation functions are shown at the
average energy of $E_{\alpha,{\rm{lab}}} = 13.5$\,MeV using the standard
potential with (full square) and the calculation with $w = 1.0$ and the
reduced imaginary radius $R_S = 1.02$\,fm 
(open square). There is almost no broadening of the derivative
$d\eta_L/dL$ at low energies which is found only for \snt
\rhes \snt\ (see Fig.~\ref{fig:eta_he6}). The data points for each $L$ are
connected by dotted lines to guide the eye. Further discussion see text.
}
\end{figure}

For a better understanding of the different behavior of the \snt \raa
\snt\ and \snt \rhes \snt\ scattering data we show in Figs.~\ref{fig:eta_he6}
and \ref{fig:eta_a} the reflexion coefficients 
$\eta_L$ which are related to the scattering matrix $S_L$ by $S_L = \eta_L
\times \exp{(2i\delta_L)}$; the reflexion coefficients $\eta_L$ and the phase
shifts $\delta_L$ are real. The shown $\eta_L$ correspond 
to the S-matrices from the calculations of Figs.~\ref{fig:scat_he6} and
\ref{fig:scat_a_low}. Both data sets show the usual behavior from $\eta_L$
close to zero for small angular momenta $L$ (corresponding to almost total
absorption), increasing $\eta_L$ for intermediate $L$ (partial absorption),
and $\eta_L \approx 1$ (no absorption) for large $L$. Again usual, with
increasing energy the number of partly or totally absorbed partial waves
increases. However, there are also significant differences in the shown
$\eta_L$ in Figs.~\ref{fig:eta_he6} and \ref{fig:eta_a}.

The slope of the $\eta_L$ {\it{vs.}}\ $L$ curves is different for the
\hes\ and the \al\ data. Therefore, we plot the slope $d\eta_L/dL$ of this
curve 
\begin{equation}
\frac{d\eta_L}{dL} := 
\frac{\eta_{L+1} - \eta_{L-1}}{(L+1) - (L-1)} = (\eta_{L+1} - \eta_{L-1})/2
\label{eq:slope}
\end{equation}
in the upper parts of Figs.~\ref{fig:eta_he6} and \ref{fig:eta_a}. One finds
curves with a shape close to Gaussian
\begin{equation}
\frac{d\eta_L}{dL} \approx
a \times \exp{\Big[-{\frac{(L-L_{0})^2}{(\Delta L)^2}}\Big]}
\label{eq:width}
\end{equation}
with the maximum slope at the angular momentum $L_0$ and the width $\Delta
L$. In general, the width $\Delta L$ is larger for the \hes\ data than for the
\al\ data. And in addition, a significant increase of the width $\Delta L$
towards lower energies is found for the \hes\ data which is not present in the
\al\ data. Significant absorption is found for all partial waves with $L \le
L_0 + \Delta L$. 

\begin{figure}
\includegraphics[bbllx=55,bblly=40,bburx=525,bbury=430,width=\columnwidth,clip=]{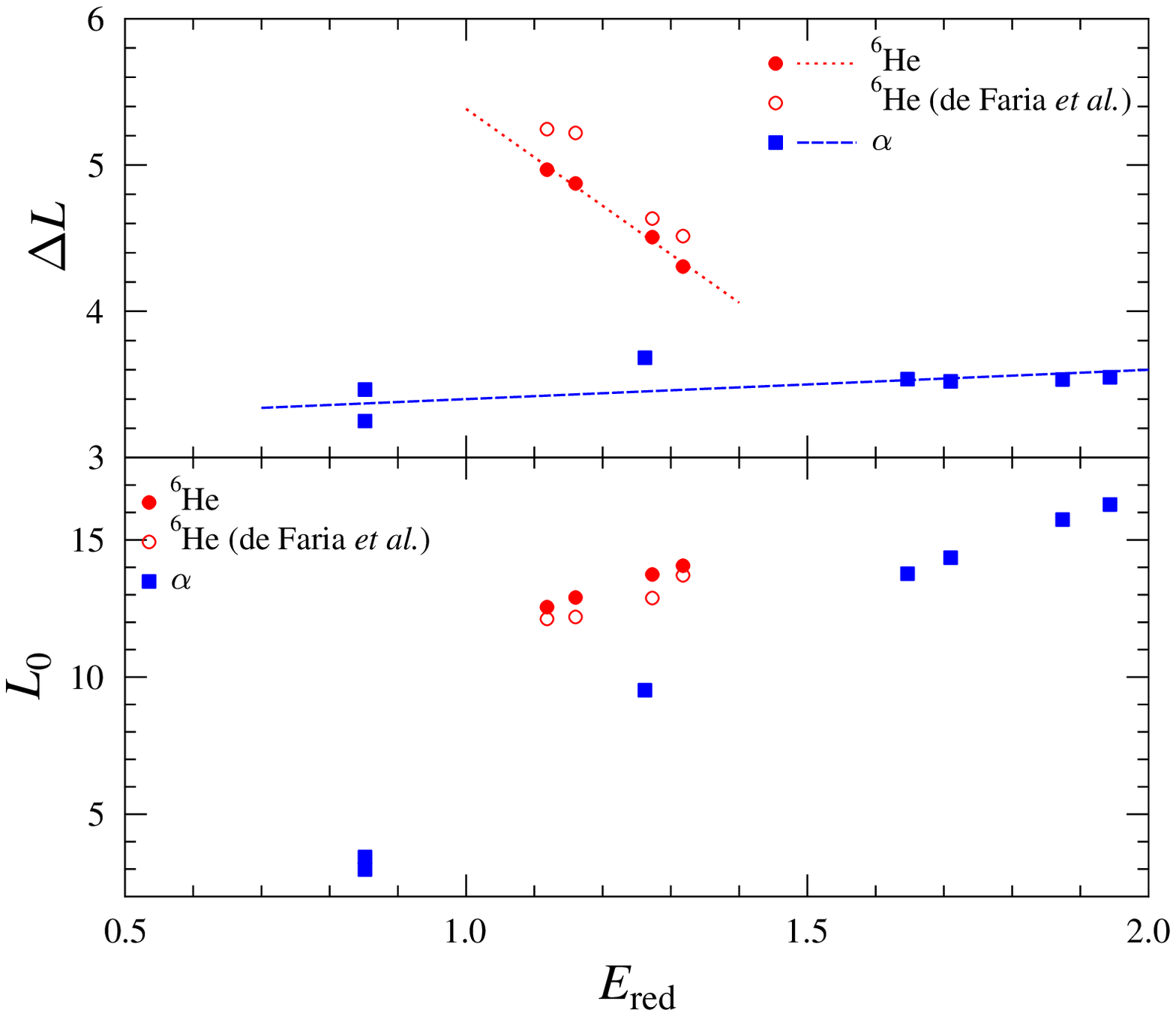}
\caption{
\label{fig:eta_shift}
(Color online)
Position $L_0$ of the maximum derivative $d\eta_L/dL$ for \snt \rhes \snt\ and
\snt \raa \snt\ elastic scattering versus the reduced energy $E_{\rm{red}}$ in
Eq.~(\ref{eq:red}) (lower part) and the Gaussian width $\Delta L$ of
$d\eta_L/dL$ in Eq.~(\ref{eq:width}) (upper part). A clear broading of
the width $\Delta L$ can only be seen for \snt \rhes \snt , whereas the width
$\Delta L$ is almost constant for \snt \raa \snt . The data for \snt \raa
\snt\ have been taken 
from the angular distributions in Fig.~\ref{fig:scat_a_low}; the two points at
the lowest energy result from the analysis of the excitation functions in
Fig.~\ref{fig:scat_a_exci} using either the standard potential or the
potential with $w = 1.0$ and the reduced imaginary radius parameter $R_S =
1.02$\,fm. The lines are to guide the eye. The open symbols show the
result of the original analysis in \cite{Far10}.
}
\end{figure}

For a better comparison of the \hes\ data and the \al\ data which have been
measured at slightly different energies, we use the so-called reduced energy
\begin{equation}
E_{\rm{red}} = E \times \frac{A_P^{1/3} + A_T^{1/3}}{Z_P Z_T}
\label{eq:red}
\end{equation}
which takes into account the Coulomb barrier (which is the same for \hes\ and
\al ) and the different sizes of the \snt -\hes\ and \snt
-\al\ systems. The obtained values for the position $L_0$ of the maximum slope
of $\eta_L$ and the width $\Delta L$ are shown in dependence of the reduced
energy $E_{\rm{red}}$ in Fig.~\ref{fig:eta_shift}.

It is obvious from Fig.~\ref{fig:eta_shift} that the maximum slope of
$d\eta_L/dL$ is found for larger $L_0$ in the \hes\ case at the same reduced
energy $E_{\rm{red}}$, thus reflecting the larger mass and momentum and the
larger absorption radius of the halo nucleus \hes . And, more important, the
width $\Delta L$ is larger for \hes\ at the same $E_{\rm{red}}$ and increases
significantly with decreasing energy. A similar effect is not seen for \snt
\raa \snt , and such a significant increase of the width $\Delta L$ is also
not found in a series of high precision \al\ scattering data in this mass
region on $^{89}$Y, $^{92}$Mo, $^{106,110,116}$Cd, and $^{112,124}$Sn
\cite{Kiss09,Ful01,Kiss06,Gal05}. These interesting findings for \hes\ 
are directly related to the energy dependence of the imaginary radius
parameter $R_S$ in the \hes\ case.

For a demonstration of the strong influence of $R_S$ in the \hes\ case we show
in Fig.~\ref{fig:eta_he6} the reflexion coefficients using the narrow
imaginary potential from 20\,MeV for the 17\,MeV data and vice versa (open
symbols); these calculations are in clear disagreement with the measured data,
see Fig.~\ref{fig:scat_he6}. The narrow 20\,MeV potential used at 17\,MeV
leads to a maximum of $d\eta_L/dL$ at lower $L_0$ and a smaller width $\Delta
L$. In parallel, $\sigma_{\rm{reac}}$ is reduced from 1479\,mb to
1114\,mb. The wide 17\,MeV potential used at 20\,MeV leads to an increased
$L_0$, a larger width $\Delta L$, and an increased $\sigma_{\rm{reac}} =
1950$\,mb instead of 1546\,mb. In the \al\ case, a similar result is found in
the calculations where the increased radius parameter $R_S$ at the lowest
energy leads to an increased $L_0$, larger width $\Delta L$, and an increased
$\sigma_{\rm{reac}} = 1459$\,mb instead of 1121\,mb. As can be seen
from Fig.~\ref{fig:scat_a_low}, the experimental data at 20\,MeV are not
reproduced using the larger radius parameter, and thus such an increase of
$L_0$ and $\Delta L$ is excluded by the new \snt \raa \snt\ scattering
data. The description of the excitation functions at lower energies requires
either a reduced diffuseness $a_S = 0.43$\,fm or a reduced radius $R_S =
1.021$\,fm in combination with $w=1.0$, but does clearly not require any
increased imaginary radius as derived from the low-energy \hes\ data. Again,
this clearly excludes any increase in $L_0$ or $\Delta L$ in the \al\ case
(see Fig.~\ref{fig:eta_shift}).

In summary, we find the following properties of the \snt -\al\ potential. The
high-energy data define the width parameter $w = 1.03$ for the folding
potential in the real part. The volume integrals $J_R$ and $J_I$ of the real
and imaginary potentials are consistent with several systematic studies. The
geometry of the imaginary part is of Woods-Saxon volume type at higher
energies; here the parameter can be fitted to the measured angular
distributions. At lower energies the surface contribution is dominating. The
imaginary diffuseness is fixed here at a standard value $a_S = 0.7$\,fm. The
reduced radius parameter $R_S$ is constant above 20\,MeV and identical to the
analysis of \snt \rhes \snt\ scattering at the highest measured energy. Only
at very low energies either $a_S$ has to be reduced, or $w = 1.0$ and a reduced
imaginary radius $R_S = 1.02$\,fm have to be used. In any case, there is no
significant broadening of the $d\eta_L/dL$ {\it{vs.}}\ $L$ curve; a
significant broadening of $d\eta_L/dL$ is only seen for the \hes\ case.

We have repeated the above analysis of the slope $d\eta_L/dL$ with the
original Woods-Saxon potentials which were fitted to the experimental \snt
\rhes \snt\ data \cite{Far10}. The same general behavior of $L_0$ and $\Delta
L$ is found from this analysis (see open symbols in
Fig.~\ref{fig:eta_shift}). Thus, it can be concluded that the experimental
\snt \rhes \snt\ data clearly require a larger value $L_0$ and an increasing
width $\Delta L$ at lower energies. This finding is independent whether
systematic folding potentials or fitted Woods-Saxon potentials are applied in
the analysis. Consequently, this increase of the width $\Delta L$ in the 
$d\eta_L/dL$ {\it{vs.}}\ $L$ curve may be taken as a signature for
the halo properties of the \hes\ projectile. Whereas $\Delta L$ changes by
about $+0.2$ for $E_{\rm{red}}$ between 0.85\,MeV and 1.85\,MeV in the \al\ case,
an one order of magnitude stronger variation of about $-0.6$ within a much
smaller range of $1.1\,{\rm{MeV}} \le E_{\rm{red}} \le 1.3\,{\rm{MeV}}$ is
found for the \hes\ case:
\begin{eqnarray}
\frac{\Delta(\Delta L)}{\Delta E_{\rm{red}}} & \approx & +0.2/{\rm{MeV}} 
\ \ \ {\rm{for}}\ \alpha \\
\frac{\Delta(\Delta L)}{\Delta E_{\rm{red}}} & \approx & -3.0/{\rm{MeV}}
\ \ \ {\rm{for}}\ ^6{\rm{He}}
\label{eq:def}
\end{eqnarray}
Halo properties may be assigned as soon as the variation of $\Delta L$ with
$E_{\rm{red}}$ is clearly below
a value of $\Delta(\Delta L)/\Delta E_{\rm{red}} \approx -1/$\,MeV around
$E_{\rm{red}} \approx 1$\,MeV.

The increase of the imaginary radius parameter $R_S$ has been explained in
\cite{Mohr00a} with the fact that the area where reactions may occur moves to
larger distances at lower energies. This has been clearly shown e.g.\ for
low-energy capture data in the $^{16}$O\rpg $^{17}$F reaction
\cite{Cho75,Mor97,Ili08}. Further work is required to follow this idea in more
detail.

The systematic behavior of the potentials in the real and imaginary parts may
be used as the basis for the construction of a simple global \hes\
potential. Because of the smooth variation of all parameters the predictive
power of such a global \hes\ potential should be very good. In particular, it
has to be pointed out that the so-called ``threshold anomaly'' is avoided in
the present study. Such ``threshold anomalies'', i.e.\ potentials with a
strong or unusual energy dependence at energies around the Coulomb barrier, or
with unusual geometry parameters like e.g.\ a huge imaginary diffuseness $a_S$
of several fm,
had to be used in many studies to reproduce the huge total reaction cross
sections of halo nuclei around the barrier (e.g.\ \cite{Far10,Agu00}). For a
deeper discussion of threshold anomalies and dynamic polarization potentials,
see e.g.\ \cite{Mah86,Tho89,Kai94,Fer07}).

For completeness, it has also to be pointed out that an unusually large
reaction cross section is not already a clear signature of a halo wave
function. Such an unusual $\sigma_{\rm{reac}}$ only indicates the strong
coupling to other channels which may not at all be related to halo
properties. E.g., such a behavior has been found in the elastic scattering of
$^{18}$O by $^{184}$W where the coupling to the low-lying $2^+$ state of
$^{184}$W leads to an unusual elastic scattering angular distribution and a
huge $\sigma_{\rm{reac}}$ \cite{Tho77,Lov77,Sat91}.

\section{Summary and conclusions}
\label{sec:sum}
We have presented new experimental data for \snt \raa \snt\ elastic scattering
at energies around and slightly above the Coulomb barrier which were measured
simultaneously with a recent \snt \rhes \snt\ experiment. The data are
successfully analyzed using systematic folding potentials in the real part and
smoothly varying Woods-Saxon potentials in the imaginary part. These
potentials are also able to reproduce \snt \raa \snt\ angular distributions at
higher energies and excitation functions at lower energies
which are available in literature. A comparison with the \snt
\rhes \snt\ scattering data shows that similar potentials with a smooth
mass and energy dependence are also able to reproduce these data. Thus, this
smoothly varying potential may be used as the basis for the construction of
simple global \hes\ potential with expected good predictive power.

The halo properties of \hes\ lead to an enhanced total reaction cross section
at low energies which is related to a relatively small elastic scattering
cross section at intermediate and backward angles. This behavior requires --~as
the only special feature for the \hes\ case~-- an energy-dependent radius
parameter $R_S$ which increases towards lower energies. Such an increase of
the radius parameter $R_S$ is not seen in the new \snt \raa \snt\ data and was
also not found in a series of high precision \al\ scattering of neighboring
target nuclei around 20\,MeV. At very low energies even an opposite trend is
seen in the analysis of the excitation functions of \cite{Tab75}.

The increase of the radius parameter $R_S$ of the \hes\ potential towards
lower energies is related to a relatively smooth rise of the reflexion
coefficients $\eta_L$ as a function of angular momentum $L$. In particular, it
is found that the width $\Delta L$ of the almost Gaussian shaped slope
$d\eta_L/dL$ is significantly larger for \hes\ compared to \al . The width
$\Delta L$ shows an increase towards lower energies for \hes\ which is not
present in the \al\ scattering data. This characteristic behavior of the
\hes\ data can be used as a signature for the halo properties of \hes , and it
should be tested as a general signature of halo properties in elastic
scattering in other cases like e.g.\ $^{11}$Be. We suggest a value below
$\Delta(\Delta L)/\Delta E_{\rm{red}} \approx -1/$\,MeV at $E_{\rm{red}}
\approx 1$\,MeV as signature for halo
properties. Although the quality of the presented new \snt \raa
\snt\ scattering data is clearly inferior compared to recent high precision
data in this mass region, only the combined analysis of the new data for \snt
\raa \snt\ scattering together with angular distributions at higher energies
and excitation functions at lower energies enables the comparison between \snt
\raa \snt\ and \snt \rhes \snt\ elastic scattering and the derivation of the
above new results.

\begin{acknowledgments}
We thank I.\ Brida and F.\ Nunes for providing their numerical results of the
\hes\ density.
This work was supported by OTKA (NN83261).
The authors wish to thank the Funda\c{c}\~ao de Amparo \`a Pesquisa do Estado 
de S\~ao Paulo (FAPESP) and the Conselho Nacional de Desenvolvimento 
Cient\'ifico e Tecnol\'ogico (CNPq) for financial support.
\end{acknowledgments}

\end{document}